\documentclass[twocolumn,showkeys,hyperref]{revtex4}
\usepackage{xspace}
\usepackage{graphics}
\usepackage{SIunits}
\usepackage{verbatim}
\usepackage{amsmath}

\newcommand{\latin}[1]{\emph{#1}}

\newcommand{\ben}{\begin{equation}}
\newcommand{\een}{\end{equation}}
\newcommand{\bean}{\begin{eqnarray}}
\newcommand{\eean}{\end{eqnarray}}
\newcommand{\be}{\begin{equation}}
\newcommand{\ee}{\end{equation}}
\newcommand{\bea}{\begin{eqnarray}}
\newcommand{\eea}{\end{eqnarray}}

\begin{document}

\title{Agent-Based Modeling of Host-Pathogen Systems: The Successes and Challenges}
\date{\today}

\author{Amy L.\ Bauer\footnote{ALB and CAAB both contributed equally to this manuscript.}}
\affiliation{Theoretical Division, Los Alamos National Laboratory, Los Alamos, NM, USA}

\author{Catherine A.A.\ Beauchemin\footnotemark[1]}
\affiliation{Department of Physics, Ryerson University, Toronto, ON, Canada\\
Center for Nonlinear Studies, Los Alamos National Laboratory, Los Alamos, NM, USA \\
Theoretical Biology and Biophysics, Los Alamos National Laboratory, Los Alamos, NM, USA}

\author{Alan S.\ Perelson\footnote{Corresponding author. Mailing address: Los Alamos National Laboratory, MS-K710, T-10, Los Alamos, NM 87545, USA. Phone: +1 505 667-6829. Fax: +1 505 665-3493. E-mail: \texttt{asp@lanl.gov}.}}
\affiliation{Theoretical Biology and Biophysics, Los Alamos National Laboratory, Los Alamos, NM, USA}

\begin{abstract}
Agent-based models have been employed to describe numerous processes in immunology. Simulations based on these types of models have been used to enhance our understanding of immunology and disease pathology. We review various agent-based models relevant to host-pathogen systems and discuss their contributions to our understanding of biological processes. We then point out some limitations and challenges of agent-based models and encourage efforts towards reproducibility and model validation. 
\end{abstract}

\keywords{ agent-based model, host-pathogen dynamics, artificial immune system, multiscale, tumor growth, tuberculosis, acute inflammation, sensitivity analysis}

\maketitle

\section{Introduction}

The ideas gleaned from studying immunology and host-pathogen systems may be relevant not only to human health but also to a wide array of other systems. A pathogen is any infectious agent that can lead to illness or disease of a host. Examples of pathogens include human immunodeficiency virus (HIV), \textit{Mycobacterium tuberculosis}, the etiological agent for tuberculosis, the SARS coronavirus, and the influenza virus. In these cases, the host is usually a human being or an animal. However, in general terms,  a host could just as easily be a computer network and the pathogen a computer virus. The immune system has memory and learns about the pathogens it encounters. It also must discriminate between self and non-self. Consequently, applications of the ideas intuited from immune system dynamics can be translated into algorithms relevant to learning, pattern discrimination, artificial intelligence adaptive behavior, and applied towards goals such as the development of new computer virus security applications \cite{forrest94,kephart95,forrest96,hofmeyr98,hofmeyr00,li05,sarafij05}. The interested readers should see \cite{forrest07}.

Because of the difficulty in reasoning about large numbers of interacting components with non-linear interactions, mathematical modeling and simulation are becoming important research tools. Depending on what aspect of the host-pathogen system is being investigated, different mathematical modeling tools are employed. Below we discuss several modeling techniques commonly used to describe such systems.

Ordinary differential equation (ODE) models are often used as a starting point to describe host-pathogen systems. One of the advantages of using ODEs is that a lot is known about their behavior. ODE models are simple and elegant and require fewer parameters than their spatial counterparts (e.g., agent-based models or partial differential equations). This is an important consideration when experimental data is obtained from a well-mixed compartment, such as blood, or from a homogenate of a tissue, such as the spleen or lymph node, where spatial information is lacking. For example, in a recent model of influenza infection in humans, Baccam et al. \cite{baccam06} used an ODE model not because the body is well-mixed but due to the fact that the data being analyzed was the amount of virus in nasal wash. When sufficiently simple, ODEs have the added benefit of being analytically tractable, and techniques such as bifurcation theory can be used to predict, for example, parameter values that switch a system from pathogen clearance to a chronically infected state. There are, however, limitations to using ODE models. These models assume that the populations (dependent variables) are homogeneous and uniformly distributed over the simulation space for all time. This is an assumption that may not be realistic, and that may significantly influence the resulting dynamics. To examine the effects of the assumption of spatial homogeneity, there is a growing body of research probing the effects of spatial distribution on systems in ecology \cite{durrett94,durrett-levin94,young01}, epidemiology \cite{hagenaars04,lloyd96}, and immunology \cite{funk05,louzoun01,strain02,cbeau06}. One option to address this limitation is to use partial differential equations (PDEs), which capture changes in both time and space, but, in general, as these equations get more complicated, and consequently more computationally challenging, the advantages to using PDE models wane. For both ODE and PDE models, one must also consider that solutions to these equations only provide an average or mean field description of the system behavior with little or no information about the possible deviations from this aggregated behavior.

An alternative to differential equation models are agent-based models (ABMs). ABMs are stochastic models used to describe populations of interacting agents, such as insects and people, using simple rules that dictate their behaviors.  These models were originally introduced by John von Neumann and Stanislaw Ulam under the name of ``cellular spaces" as a possible idealization of biological systems. They sought to show that biological processes such as the reproduction and evolution of organized forms could be modeled by simple cells following local rules \cite{hecht}. A well known feature of ABMs is their ability to generate surprisingly complex and emergent behavior from very simple rules, including periodic behaviors or intricate spatial and temporal patterns \cite{wolfcomplex}. Agent actions are asynchronous, that is, they do not evolve at constant time steps. Instead, agents respond dynamically and independently to changing environmental or discrete event cues. Consequently, nonlinearities and time-delays are not difficult to treat empirically since they can be incorporated into the agent's rules or they may even emerge naturally as a consequence of the system's collective dynamics. Another advantage to ABMs is that their computational structure is inherently parallel and therefore can be implemented on parallel computers very efficiently.

In this article, we review a variety of agent-based modeling approaches and their contributions to our understanding of host-pathogen interactions and disease dynamics.

\section{Applications of Agent-Based Models}

Agent-based models are quickly gaining in popularity. As experimental assays are developed that increase our understanding of host-pathogen interactions, the level of description desired for realistic and relevant models is increasingly more complex. Computer processing is becoming faster and more efficient, expanding the computational ability of computers and making possible the use of ABMs for complex systems. Moreover, because ABM implementation is achieved at the agent level, the description of the agents and rules tends to mimic the language used to describe the real system, that is, the description is more physical in character than mathematical. For example, models using differential equations consider rates of creation, rates of death, rates of binding, or rates of diffusion for whole populations of agents, whereas an ABM considers the rules guiding the actions of the agent. The familiar and natural modeling language used in agent based models not only makes ABMs approachable and useful to experimentalists and clinicians, but also engages them in the modeling process.

\subsection{ABMs as Immune System and Disease Simulators}

A number of immune simulators have been constructed that provide a programming framework that in principle could incorporate all current knowledge of immunology and could be used to model any aspect of immune dynamics. Platforms like IMMSIM \cite{celada92,seiden92,bezzi97,celada98,bernaschi01,celada96,klein00,kohler00,baldazzi06}, SIMMUNE \cite{meier02,meier01}, Reactive Animation \cite{efroni03,efroni05,efroni06}, and SIS \cite{cohn02,langman02,mata07} are true immune simulators and allow users to define the rules of interactions and simulate an immune reaction. Most of the immune simulators are developed to make the interaction rules simple to define and easy to change in order to facilitate the exploration and impact of different rules on the development and outcome of an immune response. Some of these simulators are more flexible than others. With IMMSIM, for example, one would typically tune interactions by changing parameter values and one could, for example, completely turn off the humoral response by setting to zero the right set of parameters \cite{kohler00}. With a system like Reactive Animation, users can potentially go even further and choose between what the authors refer to as ``running theories" \cite{efroni03}. When the rules of interactions between, for example, an epithelial cell and a T cell are not known there typically exist various hypotheses for the way in which such interactions proceed. The Reactive Animation system would allow the user to choose between different hypotheses and observe the impact of that choice on the simulation dynamics.

The usefulness and applicability of these simulators vary, but some have been applied to important immunological problems and their findings published in experimental journals. For example, IMMSIM was used to model affinity maturation and hypermutation in the humoral immune system \cite{celada96}, to test approaches to vaccine design \cite{kohler00}, and to investigate mechanisms for tolerance to pathologic rheumatoid factors \cite{stewart97}. IMMSIM has also been used as a pedagogical tool: Prof.\ Martin Weigert used IMMSIM in his seminar ``Why Immune Systems Fail: Autoimmunity, Influenza Pandemics and HIV" at Princeton to demonstrate rather than just describe immune cell interactions \cite{schultz01}. SIMMUNE became the cornerstone of the Program in Systems Immunology and Infectious Disease Modeling (PSIIM) at the National Institute of Allergy and Infectious Diseases \cite{meier-nihnews}, a program dedicated to the use of computational approaches to problems in immunology. Recently, immunologist Ronald N.\ Germain, in collaboration with SIMMUNE's creator, Martin Meier-Schellersheim, and others used the simulator to investigate the mechanism of chemo-sensing \cite{meier06}. 

Another type of simulator, which we refer to as a \emph{disease simulator}, is a general programing framework that can be tuned to model a specific disease including tumor growth, tuberculosis, or influenza.  By changing parameters, such as the rate of spread of the infection, the lifespan of infected cells, or the binding rates of cytokines, users can calibrate these frameworks to model a variety of diseases. Three such simulators are CyCells \cite{warrender-cycells,warrender-phd,warrender04,warrender06}, PathSim \cite{polys04,duca08} and the MASyV modules ma\_immune and ma\_virions \cite{masyv,cbeau06,cbeau-icaris06}. These simulators reproduce a variety of host-pathogen interactions, and are typically easier to use and calibrate than the immune simulators. 

In contrast to general disease simulators, there is another class of ABMs that concentrates instead on a single disease or on specific aspects of immune mechanisms. Among these are models of: HIV \cite{strain02,zorzenon01}, \latin{M.\ tuberculosis} \cite{segovia04,warrender06}, Epstein-Barr \cite{duca07,duca08}, influenza \cite{cbeau05,cbeau-icaris06}, cancer vaccination \cite{pappalardo05,motta05}, tumor growth and invasion \cite{mansury02,jiang05,byrne05,zhang07}, tumor-induced angiogenesis \cite{bauer07}, and acute inflammation \cite{an01,an04}. Some of these simulators have greatly contributed to improving our understanding of the dynamics of a disease and represent novel and original approaches to modeling.

\subsection{The Use of Agent-Based Models by Experimentalists}

The appeal of ABMs is such that their use goes beyond the realm of traditional modelers. While the involvement of experimentalists in mathematical modelling remains somewhat limited, ABMs in immunology/disease modeling have engaged experimentalists since their beginning. Indeed, one of the first ABM of immune interactions, IMMSIM, was the result of a close collaboration between Philip E.\ Seiden, a computer scientist, and Franco Celada, a well-respected experimentalist. This is a testament to the appeal of ABMs, where the description and representation of disease systems are close to that of the true biological system.  This attraction for experimentalists is helping to bridge the gap between theoretical models and experiments.

Marc Jenkins of the University of Minnesota is one such experimentalist. New imaging methods have, for the first time, permitted the visualization of individual T cells and their interactions with antigen-presenting cells \emph{in vivo} \cite{miller04jem,mempel04,bajenoff06,germain05}. Jenkins and his team assembled the emerging literature into a coherent picture of the first 50 hours of a primary immune response to an antigen in the form of an animation \cite{catron04}. Their model, or rather their animation, was implemented as a movie simulation using Macromedia Flash MX. It consists of a two-dimensional plane, which corresponds to a \unit{10}{\micro\metre} slice (approximately one cell diameter) through a hypothetical spherical lymph node \unit{2}{\milli\metre} in diameter, and includes T cells, B cells, and dendritic cells. Random motion paths for each cell in the simulation are pre-scripted individually and scaled to their known approximate speeds taken from the two-photon microscopy literature \cite{miller02,miller04pnas}. Restrictions are put on the paths, such that B cells are confined to the follicles (except 5 hours after exposure to antigen when their movement is restricted to the outer edge of the follicles near the T cell area), dendritic cells to the T cell area, and T cells to the T cell area and outer edges of the follicles for 90\% and 10\% of their paths, respectively. When a collision occurs between two cells, the cells' motion along their respective path is halted for a defined amount of time to mimic real-time interactions. The time alloted to each interaction depends on the circumstances. The restrictions on the cell paths and the scripted interaction times were all taken from published experimental results.

Jenkins' approach is novel; rather than writing what they feel would be a good model at an appropriate level of detail and then seeking empirical data to calibrate the model, their model was built piece by piece from the experimental literature. The resulting simulation constitutes a very complete and convincing visual summary of the available knowledge at the time and is exemplary of the capability of ABMs to translate and synthesize a large body of compartmentalized research on a complex biological system. Moreover, movies of Jenkins' simulations make other important biological contributions. For example, these movies showed that it is possible to have B-T cell interactions in the absence of directed cell chemotaxis. Indeed, random mechanisms based on restriction of B cells to a specific spatial area rarely patrolled by T cells can also result in T-B cell interactions if T cells proliferate first, thus increasing the probability of a T-B cell encounter \cite{catron04}.

Another experimentalist seduced by the appeal of ABMs is Gary An, an M.D. in Trauma and Critical Care at Northwestern University Feinberg School of Medicine. Dr. An employed agent-based modeling to describe systemic inflammatory response syndrome and multiple organ failure \cite{an01,an06}, the first application of agent-based modeling to critical care. An important aspect of this ongoing work is the utilization of agent-based modeling to simulate clinical trials, providing a crucial link between theoretical modeling and the clinical setting \cite{an04}. This model assumes that the primary immune response is initiated by endothelial cell injury at the cell-blood boundary and that the inflammatory response is translated from a local response to a systemic one. Model agents are endothelial cells, neutrophils, monocytes, several populations of T helper cells, and neutrophil/monocyte/T lymphocyte progenitor cells. The model also considers the role of a number of cell receptors, including the ICAM, IL-1, and TNF-$\alpha$ receptors, and relevant chemical mediators. Compared with results from other clinical trials and animal studies, this model reproduced the general behavior of the innate inflammatory response as measured by patient outcome and cause of death. \latin{In silico} trials of anti-cytokine therapy were simulated and compared qualitatively well with those in published phase III anti-cytokine clinical trials. Dr. An then formulated a series of hypothetical treatment strategies and used his model to evaluate the efficacy of the proposed interventions. Even this relatively simple ABM of the immune response displays complex and counter-intuitive system dynamics. Conceptually, the use of ABMs to simulate clinical trials holds a great deal of promise and sets the stage for other theoreticians and experimentalists to use models to formulate and test treatment strategies before commencing actual animal or human trials.

\subsection{Studying Localized Spatial Effects}

ABMs are stochastic models; as such they can reveal unique dynamics resulting from very specific spatial configurations or from rare localized events that would be missed using a mean field approach. This is a particularly valuable feature of these types of models, especially since similar features are responsible for variations seen in the development and outcome of infection in different individuals or in a given individual at different times. Many papers have investigated the effects of varying spatial configurations or the occurrence of rare localized events on infections \cite{cbeau06,funk05,strain02,louzoun01}.

Louzoun et al.\ \cite{louzoun01} investigated the effects of spatial heterogeneity of antigen concentration on lymphocyte proliferation. First, the authors define a simple system with two species: antigen and lymphocytes. The antigen population grows at a constant rate and decays at a rate proportional to its concentration. Lymphocytes grow at a rate proportional to both their concentration and to the concentration of antigen, and die at a rate proportional to their concentration. For this model, the authors showed that if the concentration of antigen was above some threshold, the lymphocytes would proliferate, otherwise they would not. The authors then discretized the antigen and lymphocyte populations in both space and time, and incorporated the discrete addition or removal of agents (lymphocytes or antigen) with a given probability rather than at a given rate. This change resulted in the emergence of local \emph{hot spots} of lymphocyte proliferation. So while the average antigen concentration was not above the threshold required for lymphocyte proliferation, regions where the local antigen concentration was above the threshold quickly dominated the system average with a growth rate proportional to the peaks of local concentration of antigen, and not to the average concentration throughout the system \cite{louzoun01}.

Strain et al. \cite{strain02} examined the specific case of an HIV infection and compared the behavior of a commonly used ODE model \cite{perelson96hiv,perelson97hiv} to an analogous cellular automaton model where each lattice site represents a T cell. From the diffusion rate of virions and the rate of encounter and fusion between virus and T cells, the authors derived an expression for the probability that a T cell located a certain distance away from a bursting infected T cell will become infected. The authors initiate their simulation with a single infected cell which in turn infects other cells according to the probability they derived. Subsequent infections result in a radial wave of infection moving outward. This wave leaves behind a pool of infected cells. If the replacement rate of dead cells with new target cells is slow, the wave dissipates and uninfected cells replenish the area. If the replacement rate is fast, the newly regenerated target cells cause the infection wave to recede and a chaotic steady state pattern forms \cite{strain02}. Most importantly, because infectivity depends on the concentration of T cells (neighbors can be infected more readily when cells are tightly packed), the authors found that the ODE model overestimated viral infectiousness by more than an order of magnitude when compared with the spatial model \cite{strain02}. 

Funk et al.\ \cite{funk05} used an approach similar to that of Louzoun et al.\ \cite{louzoun01}, but proposed a more complete and realistic model of three species: uninfected target cells, virus-infected cells, and virus. The authors first present the dynamics of the non-spatial homogeneous model, and then extend the model to a 2-D spatial model, and finally they randomly vary the parameters from site to site according to a uniform distribution. Comparing results from the non-spatial homogeneous model with the spatial model, the authors found that as the infection spreads outwards from the initial infection site, the viral titer curves for both models are initially similar. However, as the initially infected sites approach an equilibrium state and outer cells become affected by the infection wave, the increase in virus slows. This phenomenon is not seen in the non-spatial homogeneous model. As in the Strain et al.\ model, the Funk et al.\ spatial models also resulted in lower average viral concentrations when compared to non-spatial homogeneous models \cite{funk05}.

When the parameters of the Funk et al.\ model were varied from site to site, a new dynamic was revealed. This new system formed a series of source and sink sites and the ability of the virus to migrate from site to site (its diffusion rate) had a significant impact on the dynamics. Finally, the authors added immune cells to their heterogeneous spatial model. For this case, they found that a spatial model greatly improved the stability of the infection. Stability was enhanced because the equivalent non-spatial homogeneous model gave rise to large oscillations, analogous to what is observed in predator-prey models in ecology. In contrast, the spatial coupling by local dispersal of virus and immune cells in the non-homogeneous spatial model resulted in oscillations at different sites that were out of phase. These oscillations were damped and had the effect of spatially averaging the dynamics. This is a very interesting finding as the large long-lasting periods of oscillations commonly observed in non-spatial models are only very rarely observed experimentally or clinically, while the equalized dynamics of spatial models are more in line with experimental and clinical observations. 

\begin{figure*}
\begin{center}
\resizebox{0.7\linewidth}{!}{\includegraphics{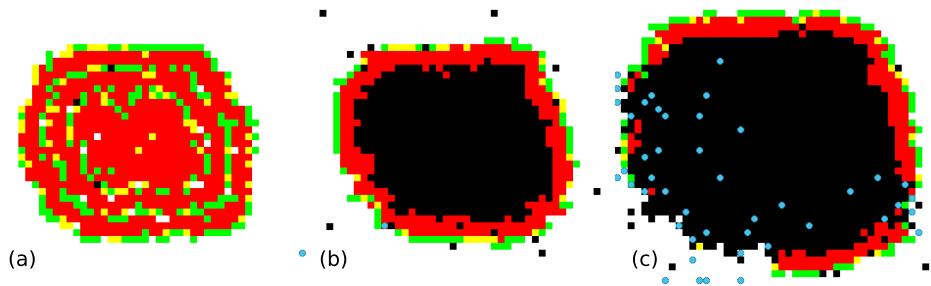}}
\end{center}
\caption{Effect of the regeneration rule on the infection dynamics. From left to right: (a) using a global rule where dead cells are replaced at a rate proportional to the total number of uninfected cells; (b) using a local rule where dead cells are only replaced when an immediate uninfected neighbor is dividing; (c) when using the local rule, dead cells can only be regenerated once the infection wave has been breached by the immune cells. Using the same infection and death rate, a local regeneration rule results in a larger number of dead cells but a smaller number of infected cells compared to a non-spatial global rule. The images are screenshots taken from MASyV's \texttt{ma\_immune} client \cite{masyv,cbeau06} and represent a 2-D tissue patrolled by immune cells (blue) where each lattice site corresponds to a tissue cell which can either be uninfected (white), dead (black), or in various stages of infection (green, yellow, red).}
\label{regen-sshot}
\end{figure*}

Finally, Beauchemin \cite{cbeau06} investigated the effect of viral infection spread in tissue and the particular effect on influenza dynamics. The model is a 2-D square lattice where each site represents a target cell and the grid, which represents the tissue, is patrolled by generic immune cells which can kill infected cells. While the models mentioned above all initiate infection with a single infected cell, this work explored the effect of altering the initial distribution of infected cells. The cases examined included a given number of initial cells distributed either randomly on the grid, in small isolated groups of several cells, or as a single lump. Beauchemin found that infection dynamics was sensitive to the initial spatial distribution. This is due to the fact that when an infected cell is part of an infected patch, it already has infected neighbors and therefore has fewer infectable neighbors than a single infected cell whose neighbors are all infectable. This results in a smaller effective infection rate in simulations where initially infected cells are distributed in larger lumps. The author then explored the effect of target cell replenishment. In ODE models, target cells are typically replenished at a fixed rate or at a rate proportional to the number of target cells as a function of cell division. Beauchemin explored the effects of changing the replacement rule such that an empty site can only be replenished by a new target cell if it is in contact with a dividing uninfected target cell. Screenshots resulting from both regeneration rules are presented in Figure \ref{regen-sshot}.

As the infection wave spread outward, it left behind a pool of dead cells, which could only be replenished if in contact with uninfected cells. This could only happen once immune cells had detected the infection and began breaching the infection wave that segregates the dead cells from the uninfected cells (see Figure \ref{regen-sshot}). This modification to the model yielded a curve for the fraction of dead cells over time which closely resembled experimentally observed curves \cite{cbeau06,bocharov94} and contrasted with the sharp and narrow curves obtained with a model where replacement rate only depends on the number of uninfected cells but not on their location.

\subsection{Infection of Cell Layers}

Because emerging spatial patterns can have a significant effect on the development and outcome of a simulated infection, it is important to define the model in a manner that is consistent with the true spatial nature of the real system. For example, two existing ABMs of HIV \cite{zorzenon01,strain02} are implemented on 2-D square lattices where each site represents a T cell and the infection spreads from infected cells to their neighbors. The infection spreads as a wave and under certain conditions, intricate wave patterns can form. Unfortunately, while it is true that T cells are tightly packed (as in a lattice) \cite{bajenoff06}, they are quite motile with velocities of approximately $\unit{11}{\micro\meter/\minute}$ \cite{mempel04,miller03}. It is likely that if the T cells in these HIV models were motile, these intricate infection wave patterns would not form.

There are, however, models where a 2-D lattice of static cells is the best and most accurate representation. An example of this are \latin{in vitro} experiments conducted on 2-D cell monolayers \cite{duca01,lam05,howat06}. Howat et al.\ \cite{howat06} used data from 2-D Madine-Darby bovine kidney (MDBK) and Vero cell cultures infected with Herpes simplex virus 1 (HSV-1) to study the antiviral effect of interferon $\beta$ (IFN-$\beta$). Their ABM consists of a 2-D hexagonal lattice where each site represents a target cell that can either be susceptible, infected, in an antiviral state, or dead. Because IFN-$\beta$ diffusion is very fast, it is taken to be instantaneous and IFN-$\beta$ is tracked as a single global variable. Because the experimental cellular monolayer is overlaid with an agar that impedes viral diffusion, infection is assumed to spread locally: Virions are released by an infected cell as a single burst upon cell death and distributed to the 6 immediate neighbors and 12 next neighbors weighted according to distance. By comparing the fraction of dead cells over time obtained experimentally to what their ABM produces, the authors were able to calibrate their model and determine key parameters of HSV-1 infection and IFN-$\beta$ protection. For example, they found that it takes about 95 hours for the MDBK monolayer to reach an antiviral state \cite{howat06}. Once calibrated, Howat et al.\ used their model to determine the dosage of HSV-1 needed to obtain total monolayer destruction, maximum IFN-$\beta$ production, or minimum recovery time. Notably, the authors determined that the minimum recovery time is obtained for the maximum boundary length, i.e., when the dead cell plaque geometry is such that it maximizes the number of cells in contact with an empty area (a pool of dead cells).

Beyond \textit{in vitro} experimental systems, there are also \textit{in vivo} diseases which can be accurately represented by a 2-D lattice of non-motile cells. An uncomplicated influenza infection, i.e., one which infects only the upper lung and does not degenerate to pneumonia, is a good example. Beauchemin et al.\ \cite{cbeau05,cbeau06} have taken advantage of this fact and developed MASyV's \texttt{ma\_immune}, a 2-D square lattice model of influenza A infection \cite{masyv}. This representation is a reasonable description of the true system because the target of influenza A is the tightly packed monolayer of cells that line the airways. Additionally, since influenza virions only bud apically (from the top of the infected cells) \cite{mora02} the infection effectively occurs in a 2-D plane. Interestingly, when this model was published in 2005, it was only the second model ever proposed for influenza infection in humans; the other one being the 60 parameter ODE model with delays proposed by Bocharov and Romanyukha 20 years earlier \cite{bocharov94}. At the time, very little experimental data on the dynamics of influenza was available and a spatial model was especially difficult to justify. Nevertheless, Beauchemin et al.\ were able to fix 3 of their parameters and set realistic biological bounds on 6 of the 8 other parameters from information gathered in the literature. Then, picking parameter values within these bounds, the authors were able to reproduce a number of features of influenza infection, for example, that the peak of infected cells occurs at \unit{48}{\hour} and that infection resolves by day $5\pm2$. This work was originally intended to show that the ABM is complex enough to reproduce the general shape of a response to an uncomplicated viral infection, and that it yields quantitatively reasonable results when parameterized for a particular viral infection. The ABM was later used to study the influence of spatial heterogeneities on the dynamical evolution of a viral infection \cite{cbeau06}, as discussed in Section 2.3.

This model also improved our understanding of flu kinetics. The model was used to derive values for some parameters (e.g.\ replication rate of cytotoxic T lymphocytes per infected cell encounter) that were subsequently employed in other influenza models \cite{chang07}. The model's agreement with the limited available data motivated its expansion into a more detailed and accurate model of influenza infection in vitro.  In collaboration with experimentalist Dr.\ Frederick T.\ Koster, M.D., and University of New Mexico computer scientist Stephanie Forrest, Beauchemin and others developed an improved version of the model, MASyV's \texttt{ma\_virions}, which accounts for viral release, dispersion, and clearance \cite{cbeau-icaris06,masyv}. The ABM consists of a 2-D hexagonal grid representing lung tissue and on which viral concentrations evolve based on a discretized version of the diffusion equation. Calibration of the ABM was accomplished by quantitatively comparing the ABM's results with experimental data for fraction of infected cells, patchiness, and viral concentrations versus time produced by an in vitro human lung cell monolayer \cite{cbeau-icaris06}. The calibration to preliminary data provided an estimate for the diffusion rate of virions that was 1,000-fold lower than the natural diffusion expected from the Stokes-Einstein equation. The calibrated virion diffusion rate is consistent with virions getting trapped by cell receptors and mucus \cite{cbeau-icaris06}.

\subsection{Agent-Based Models in Shape Space}

The space in which ABMs evolve does not always need to correspond to 3-D Euclidean space. Another variety of ABMs of host-pathogen/immune systems are called \emph{shape space} models \cite{perelson79,perelson91,stauffer92,dasgupta92,perelson97rev,smith98,stromberg06,lee07}. In these ABMs, the spatial dimensions represent the generalized shapes \cite{perelson79} of cell receptors and epitopes, the parts of antigens recognized by the immune system. The affinity between a receptor and an epitope is a function of their distance in shape space \cite{perelson79}. The evolution of the system in space and time then can be used to represent the development of an immune response including receptor variability due to somatic hypermutation, and the dynamics of clonal expansion and contraction. Typically, an epitope corresponds to a given location in shape space and its presence stimulates the activation and proliferation of lymphocytes located at and around that site, that is, those with a high receptor affinity for that epitope. In modeling viral infections it is also possible to allow the virus to mutate and evolve its epitopes so as to avoid immune elimination.  Thus, the shape space framework is very flexible and has obvious use in immunological modeling.

One shape space model that has had an impact on influenza research was proposed by Smith et al.\ \cite{smith99} in order to study the efficacy of repeated annual influenza vaccination. Although annual influenza vaccination is recommended for high risk groups, such as infants and the elderly, clinical studies of the efficacy of annual vaccination have lead to conflicting results. One empirical study by Hoskins et al. \cite{hoskins79} suggested that during some flu seasons patients given repeated annual vaccinations had less protection that first-time vaccinees, while another study suggested that repeated annual vaccination did in fact offer long-term continual protection \cite{keitel97}. Smith et al.\ \cite{smith99} showed that differences in the antigenic distance (i.e.\ the distance in shape space) between the vaccine strain and the circulating influenza strain in both studies, which were conducted in different years, could account for the observed discrepancies in the results. In particular, they showed that a first vaccine will negatively interfere with the protection potentially afforded by a second vaccine when the two vaccine strains are close in shape space. This is because the first vaccination raises antibodies that by cross-reactivity can partially eliminate the second vaccine.  Using an ABM that followed the dynamics of a broad array of B cells and antibodies for the vaccine epitopes, Smith et al.\ \cite{smith99} were able to show in a quantitative manner how the model could explain the available experimental data on the efficacy of repeated vaccination.  Further, the idea of monitoring the antigenic distance in shape space between flu strains has proved to be a valuable approach in studying influenza evolution \cite{smith04} and is now a factor in the World Health Organization's biannual decision making process for selecting which vaccine strains to include in the seasonal influenza vaccine \cite{smith06}.

\subsection{The Role of Agent-Based Models in Multiscale Systems}

Biological systems naturally span multiple time and length scale hierarchies. Time and spatial scales can range from $10^{-2}$ seconds and $10^{-9}$ meters during gene transcription and protein production at the molecular level to $10^6$ seconds and up to several meters at the level of the whole organism \cite{hunter02,segovia04}. However, experimental investigations and the data generated focus on specific mechanisms relegated to isolated time or length scales and must be translated between biological scales. A multiscale approach to modeling biological systems is crucial to developing realistic and relevant models capable of predicting complex biological phenomena.

Bridging multiple scales is a formidable modeling challenge. Because ABMs are constructed by considering the behaviors of individual system components and can be developed for each biological subsystem or hierarchy, they are also naturally well suited for linking these different models together.  For example, merging ABMs focusing on different biological scales has been used to translate the results of and facilitate collaboration between experimental groups working on various aspects of the acute inflammatory response \cite{an06}. In this work, a group of agent based models were developed to simulate intracellular signal transduction pathways that were then incorporated into ABMs for different cell types. Subsequently, these cellular models were integrated to simulate tissue and whole organ function during the acute inflammatory response. 

Another modeling technique that has emerged in response to the computational challenge presented by multiple scales integrates differential equation models with ABMs to couple the dynamics occurring on different time and length scales \cite{peirce04,byrne05,jiang05,bauer07,kirschner07,zhang07}. This technique has been applied to the development of models describing tumor growth, a complex biological system with important clinical applications \cite{byrne05,jiang05,zhang07}. In Jiang et al. \cite{jiang05}, a three-dimensional model of avascular tumor growth was developed that spans three distinct scales. A type of ABM, called a lattice Monte Carlo model, is used to describe tumor cell growth, adhesion, and viability. These cell dynamics are regulated at the intracellular level by a Boolean network for protein expression that controls the cell cycle. At the extracellular level, nutrients, metabolic waste, and growth factor and inhibitory chemical concentrations are described by a system of partial differential equations. Growth curve measurements and measurements of the size of the proliferating rim and necrotic core regions from simulated tumors agree with the quantitative results obtained from tumor spheroid experiments.  This model predicts what environmental conditions tumor cells require for survival, and the molecular weight of potential growth promoters and inhibitors regulating tumor cell viability.  As demonstrated by these results, validated models of biological systems provide new biological insights, can be used for prediction, and guide future experimental pursuits. Looking forward, coupling tumor growth models with agent-based and cell-based models of angiogenesis \cite{bauer07,peirce04} and the immune system response against the tumor will be a major step towards the development of an integrated systems approach to modeling tumor growth and cancer invasion and towards the ultimate goal of predicting the effects of novel cancer and anti-angiogenic therapies.

\subsection{Improved Experimental Data Fueling Advances in Modeling}

Models of any type can only be as good as the data used to develop, calibrate and confirm them. Fortunately, new experimental techniques are allowing access to unprecedented amounts of data, but also to novel types of data. Because ABMs are spatial models they require spatial data: a rare commodity. High resolution spatial data is especially important given that localized spatial effects can have an important impact on the resulting dynamics, as mentioned above. The recent application of two-photon microscopy to \latin{in situ} and \latin{in vivo} imaging of mice lymph nodes has determined trajectories of individual T cells as they move within lymph nodes \cite{miller02,miller03,miller04jem,germain05,bajenoff06}. This new technique has revealed that T cell motion in lymph nodes occurs largely as a result of the cells crawling along fibers of the fibroblastic reticular cell network \cite{bajenoff06}. Additionally, examination of these trajectories has shown that the end-to-end distance that T cells travel increases approximately linearly with the square root of time, suggesting that the movement of T cell over long distances can be described by a random walk rather than motility guided by chemotaxis. These new data have facilitated the elaboration of models for the movement of T cells within lymph nodes: some simple \cite{cbeau07tmvt} and some more detailed \cite{beltman07,meyer05}.

The model by Beauchemin et al.\ \cite{cbeau07tmvt} was a simple model where a T cell moves in a straight line at constant speed $v_\mathrm{free}$ for a time $t_\mathrm{free}$, then pauses for a time $t_\mathrm{pause}$ as it reorganizes its intracellular machinery allowing it to turn. It then picks a new direction, and undergoes another free run. Comparing simulations against experimental data yielded a mean free path of approximately $\unit{38}{\micro\meter}$, which is about twice the distance between intersections in the fibroblastic reticular cell network, which averages $\unit{17\pm7}{\micro\meter}$. From this, the authors concluded that when coming to a fiber intersection, a T cell will turn roughly $50\%$ of the time, and $50\%$ of the time it will continue along its original fiber.

The Meyer-Hermann and Maini model \cite{meyer05} offers a more detailed description of the active crawling exhibited by both B and T cells. This model is a 2-D cellular Potts model \cite{graner92,glazier93} where a B or T cell is represented by a group of neighboring lattice sites. The lattice sites that make up the lymphocyte are moved individually but their movements are constrained to maintain a constant cell volume and to respect the general direction chosen for active movement. The model yielded a distribution for the velocities of B and T cells that agreed impressively well with velocities obtained experimentally \cite{meyer05}.

More recently, Beltman et al.\ \cite{beltman07} proposed a 3-D rather than a 2-D cellular Potts model, which also represented lymphocytes as a collection of neighboring lattice sites. The authors investigated the possibility that motility patterns observed for T cells and dendritic cells are a result of the anatomical structure of lymph nodes, rather than an intrinsic characteristic of the cells' motility. Indeed, they found that by restraining T cell movement using the constraints imposed by anatomical barriers in lymph nodes, the model can account for all characteristics of the experimental observations. In particular, they showed that lymph node anatomy can explain the large velocity fluctuations seen experimentally for T cells. Further, they found that as a result of the obstacles, T cells tend to organize into small, highly dynamic streams and provide experimental evidence to support these new findings. Finally, using their model, they were able to predict that dendritic cells would be able to contact approximately 2,000 different T cells per hour while T cells could contact roughly 100 dendritic cells per hour \cite{beltman07}.

\subsection{Sensitivity and Uncertainty Analysis in Modeling and Prediction}

Because ABMs consist of several thousand lines of code, it can often be just as complicated to understand the model as it is to understand the real system. In general, as the system grows in complexity, the number of parameters that must be estimated also increases. Parameter estimation introduces uncertainty into the system due to experimental error, differences in experimental assays, or error from data fitting techniques.  Moreover, when the number of agents is large and their interactions numerous and complicated, it can become difficult to extract or isolate the key processes responsible for a given outcome. In their paper proposing an ABM of \latin{M.\ tuberculosis} infection, Segovia-Juarez, Ganguli, and Kirschner \cite{segovia04} use a simple and efficient way of analyzing their results by performing a sensitivity and uncertainty analysis of the model's parameters. The authors picked 12 of the 27 parameters of their model and investigated their effect on the growth of tuberculosis granulomas. Using the number of extracellular bacterium and the granuloma size as outcome variables, they mapped out the parameter space using the Latin hypercube sampling method and quantified the sensitivity of their model to these parameters using partial rank correlation coefficients on each parameter set. This type of analysis quantifies the importance and impact of each parameter on the outcome variables. Their analysis revealed that the intracellular growth rate of the bacteria is strongly and positively correlated with the number of extracellular bacteria at early times of the infection and at much later times post-infection, and negatively correlated at intermediate times. This suggests that at low bacterial levels, a large intracellular bacteria growth rate is critical to infection, while a small growth rate seems necessary to generate larger lasting granulomas. This is one example of how simulation and analysis can be coupled to provide insights into and enhance our understanding of the dynamics of disease (see \cite{blower94,kirschner08} for a more complete review of these analytic techniques).

\section{Discussion}

Agent-based models have inspired significant interest through their visual appeal and because the languages used to describe the model and the natural system are very similar. Computational advances are making possible the use of ABMs to describe whole systems arising not only in the human immune system, which has been the focus of this article, but also in financial markets \cite{hoffmann07,farmer01,vandenbergh02,farmer07}, the spread of epidemics \cite{dibble04,mueller04}, cancer dynamics \cite{jiang05,byrne05,mansury02,zhang07}, and the threat of bio-warfare \cite{carley06} to name just a few. In addition, ABMs have engaged experimentalists and have helped facilitate their increasing involvement in modeling. Here, we have presented a limited survey of ABMs in the context of host-pathogen dynamics. This survey is in no way exhaustive, but is meant to highlight the novel and original modeling approaches applied by researchers and the contributions to our knowledge of specific diseases and immune mechanisms.

ABMs are only one of a number of modeling techniques that could be applied to describe a particular system. The choice of model to employ is guided by multiple factors including the questions that one wishes to answer, the assumptions that can be made, parameter availability, and computational expense. Choosing agent-based modeling is not without cost or consequence. ABMs tend to require a larger number of parameters than both ODE models and their spatially-continuous partial differential equation analogs. Until experimental assays capable of measuring relevant experimental data catch up with the needs and complexity of model calibration, ODEs are sometimes the only viable modeling option. An important weakness of ABMs arises in the context of sharing and comparing results. At worst, a partial differential equation model with delays will consist of a large and complicated set of equations, but that can be written down in an unambiguous manner. In contrast, an ABM consists of many lines of codes which could be in any variety of programming languages, and its translation into words can overlook very important details of its implementation that are crucial to reproducing its results. This is a problem that researchers utilizing ABMs need to pay more attention to. Establishing a permanent home for any computer model developed and a static version corresponding to the exact state of the program used for published results are important first steps. Software repositories like SourceForge.net (\url{http://sourceforge.net}) and freshmeat (\url{http://freshmeat.net}) provide such services free of cost.

As with any modeling approach, ABMs need to adhere to the principles of scientific rigor. For an ABM to be useful to understanding the mechanisms of the system, it needs to convincingly relate to it. Models need to be calibrated and researchers should confirm that their parameter values are within physically relevant ranges. Toward this end, it is helpful to provide a sensitivity analysis of the parameters to identify key parameters and to fully grasp the consequences of parameter uncertainty and variability on the observed outcomes. Modelers should also make every effort possible to quantitatively validate the results of their simulations with independent experiments or with reports in the literature, using data different from those that were used for model calibration. Finally, and perhaps most importantly, models should strive to make experimentally verifiable predictions. Once verified, such predictions will establish that the model realistically captures some aspects of the system. Further, models validated in this way may offer new perspectives, provide a framework for formulating and testing hypotheses, and may suggest important new experiments.

\begin{acknowledgments}
Portions of this work were done under the auspices of the U.S.\ Department of Energy under contract DE-AC52-06NA25396 and supported by NIH grants AI28433, RR06555, P01-AI071195, and N01-AI50020 (ASP), and the UNM/LANL Joint Science and Technology Laboratory and NIH grant R21-AI73607 (CAAB).
\end{acknowledgments}

\addcontentsline{toc}{section}{References}
\bibliographystyle{abbrv}
\bibliography{allbibliographies,abm-review}

\end{document}